\documentclass[12pt,thmsa]{article}
\usepackage{amsfonts}

\usepackage{sw20aip}

\input tcilatex

\begin{document}

\title{On the path integration for the potential barrier $V_0\cosh ^{-2}(\omega x)$}
\author{L. Guechi \\
%EndAName
D\'epartement de Physique, \and Facult\'e des Sciences, Universit\'e
Mentouri, \and Route d'Ain El Bey, Constantine, Alg\'eria. \and T. F.
Hammann \\
%EndAName
Laboratoire de Math\'ematiques, Physique math\'ematique, \and Facult\'e des
Sciences et Techniques, \and Universit\'e de Haute Alsace, \and 4, rue des
fr\`eres Lumi\`ere, F-68093 Mulhouse, France}
\maketitle

\begin{abstract}
The propagator associated to the potential barrier $V=\cosh ^{-2}(\omega x)$
is obtained by solving path integrals. The method of delta functionals based
on canonical and other transformations is used to reduce the path integral
for this potential into a path integral for the Morse potential problem. The
dimensional extension technique is seen to be essential for performing the
multiple integral representation of the propagator. The correctly normalized
scattering wave functions and the scattering function are derived. To test
the method employed, the free particle and the $\delta -$function barrier
are considered as limiting cases.



typescript using Latex
\end{abstract}

\section{Introduction}

During the last twenty years, a number of nontrivial path integrals of
nonrelativistic quantum mechanical problems have been solved \cite
{khan,klei,inokurger,grostei} . In particular, two wide classes for which
the path integral solutions are related to the radial harmonic oscillator
and the P\"oschl-Teller path integral, are extensively studied. The first
class called class of confluent potentials includes the radial harmonic
oscillator, the Morse potential, and the Coulomb potential. The
P\"oschl-Teller and the modified P\"oschl-Teller potentials, e.g. the
Rosen-Morse, Manning-Rosen, Scarf-like, or the Hulthen potential belong to
the class of hypergeometric potentials. However, most of discussions of
these potentials by path integration were restricted to bound state problem.
For the second class, when the extension to describe the scattering states
was made, the approach applied is based on a Sommerfeld-Watson
transformation which leads to a closed form expression for the Green's
function \cite{durklei,kleimus,chegueleha1,chegueleha2,cheguelehame,gros1}
or via the path integration over the $SU(1,1)$ manifold for the calculation
of the propagator with bound and scattering states found simultaneously\cite
{junboh,bohjun,barinojun,bouchegueha,gros2} . Note that the noncompact group 
$SU(1,1)$ can be viewed as analytical continuation of the compact group $%
SU(2)$.

The purpose of this paper is to solve the problem of the potential barrier $%
V=\cosh ^{-2}(\omega x)$ in the framework of Feynman's path integral. This
potential is not a special case of the modified P\"oschl-Teller potentials.
It cannot be related to them by coordinate transformation and it has not a
dynamical $SU(1,1)$ symmetry. Thus, neither the approach which consists in
summing the spectral representations of the Green's function via a
Sommerfeld-Watson transformation nor the path integration over the $SU(1,1)$
manifold can be applied to this potential. A treatment of this potential
barrier has been presented recently in the Feynman approach through
integration on the $SO(1,2)$ group\cite{ahmedu} , but it involved several
incorrect manipulations so that the amplitudes $T(k)$ and $R(k)$ of the
transmitted and reflected beams obtained are not correct. We think that it
is worthwile to present another path integral approach to solve this
potential barrier problem.

In sec.II, we first introduce an auxiliary dynamics by extension of the
phase space in order to convert the path integral for the potential barrier
into a path integral form analogous to the rigid rotator problem. Then, with
the help of a point canonical transformation and the trick called
''uncompleting the square''\cite{andand} , the propagator is connected to
the path integral for the Morse potential problem. In sec.III, the
propagator is evaluated, from which we obtain the suitably normalized wave
functions. The scattering function is determined from the asymptotic form of
the wave functions in sec.IV. As a test of our calculation, two limiting
cases are considered. The coefficients of the transmitted and reflected
beams are found again for the free particle and for a particle subjected to
the repulsive $\delta -$function potential in sec.V. The section VI will be
a conclusion.

\section{Reduction of cosh$^{-2}$($\protect\omega x$) into a form of Morse
Potential}

The path integral in phase space for a particle of mass m subjected to the
potential barrier defined by

\begin{equation}
V(x)=\frac{V_0}{\cosh ^2(\omega x)};V_0\succ 0,  \label{b.1}
\end{equation}

is written formally as follows:

\begin{equation}
K(x_f,x_i;T)=\int DxDp_x\exp \left\{ \frac i\hbar \int_0^T\left( p_x%
\stackrel{.}{x}-\frac{p_x^2}{2m}-\frac{V_0}{\cosh ^2(\omega x)}\right)
dt\right\} .  \label{b.2}
\end{equation}
In the time sliced representation, it is given by\ \ 
\begin{eqnarray}
K(x_f,x_i;T) &=&\stackunder{N\rightarrow \infty }{\lim }\int \stackrel{N}{%
\stackunder{j=1}{\prod }}dx_j\stackrel{N+1}{\stackunder{j=1}{\prod }}\frac{%
dp_{x_j}}{2\pi \hbar }\exp \left\{ \frac i\hbar \stackrel{N+1}{\stackunder{%
j=1}{\sum }}\left[ p_{x_j}\triangle x_j-\right. \right.  \nonumber \\
&&\left. \left. \left( \frac{p_{x_j}^2}{2m}+\frac{V_0}{\cosh ^2(\omega x_j)}%
\right) \varepsilon \right] \right\} ,  \label{b.3}
\end{eqnarray}
where we have adopted the standard notation:

$\varepsilon =t_j-t_{j-1}=\frac T{N+1},\triangle x_j=x_j-x_{j-1},x_i=x(0)$
and $x_f=x(T)=x(t_{N+1}).$

Let us first rescale $(x,p_x)$ by $(\alpha =\omega x,p_\alpha =\frac 1\omega
p_x)$. Under this transformation, the propagator (\ref{b.3}) becomes 
\begin{eqnarray}
K(x_f,x_i;T) &=&\omega \stackunder{N\rightarrow \infty }{\lim }\int 
\stackrel{N}{\stackunder{j=1}{\prod }}d\alpha _j\stackrel{N+1}{\stackunder{%
j=1}{\prod }}\frac{dp_{\alpha _j}}{2\pi \hbar }\exp \left\{ \frac i\hbar 
\stackrel{N+1}{\stackunder{j=1}{\sum }}\left[ p_{\alpha _j}\triangle \alpha
_j\right. \right.  \nonumber \\
&&\left. \left. -\left( \frac{p_{\alpha _j}^2}{2m}+\frac{V_0/\omega ^2}{%
\cosh ^2\alpha _j}\right) \omega ^2\varepsilon \right] \right\}  \nonumber \\
&=&\omega \int D\alpha Dp_\alpha \exp \left\{ \frac i\hbar \int_0^{\omega
^2T}\left[ p_\alpha \stackrel{.}{\alpha }\right. \right.  \nonumber
\label{b.4} \\
&&\left. \left. -\left( \frac{p_\alpha ^2}{2m}+\frac{V_0/\omega ^2}{\cosh
^2\alpha }\right) \right] dt\right\} .  \label{b.4}
\end{eqnarray}

In analogy with the problem of a free particle moving on the two-\newline
dimensional sphere which is also known as the rigid rotator problem \cite
{peakinolanino} , the path integration of the expression (\ref{b.4}) can be
simplified by introducing an additional dynamics by extension of the phase
space with the help of the following path integral identity:\ 
\begin{eqnarray}
&&\ \ \exp \left[ -\frac{i}{\hbar }\left( \int_{0}^{\omega ^{2}T}\frac{%
V_{0}/\omega ^{2}}{\cosh ^{2}\alpha }dt\right) \right]   \nonumber \\
\  &=&\int_{-\infty }^{\infty }d(\beta _{f}-\beta _{i})\exp \left[ -i\sqrt{%
\frac{2mV_{0}}{\hbar ^{2}\omega ^{2}}-\frac{1}{4}}(\beta _{f}-\beta _{i})%
\right] \stackunder{N\rightarrow \infty }{\lim }\int \stackrel{N}{%
\stackunder{j=1}{\prod }}d\beta _{j}  \nonumber \\
&&\ \ \times \stackrel{N+1}{\stackunder{j=1}{\prod }}\frac{dp_{\beta _{j}}}{%
2\pi \hbar }\exp \left\{ \frac{i}{\hbar }\stackrel{N+1}{\stackunder{j=1}{%
\sum }}\left[ p_{\beta _{j}}\triangle \beta _{j}-\frac{p_{\beta _{j}}^{2}+%
\frac{\hbar ^{2}}{4}}{2m\cosh ^{2}\alpha _{j}}\omega ^{2}\varepsilon \right]
\right\}   \nonumber \\
\  &=&\int_{-\infty }^{\infty }d(\beta _{f}-\beta _{i})\exp \left[ -i\sqrt{%
\frac{2mV_{0}}{\hbar ^{2}\omega ^{2}}-\frac{1}{4}}(\beta _{f}-\beta _{i})%
\right]   \nonumber  \label{b.5} \\
&&\ \ \times \int D\beta Dp_{\beta }\exp \left\{ \frac{i}{\hbar }%
\int_{0}^{\omega ^{2}T}\left[ p_{\beta }\stackrel{.}{\beta }-\frac{p_{\beta
}^{2}+\frac{\hbar ^{2}}{4}}{2m\cosh ^{2}\alpha }\right] dt\right\} .
\label{b.5}
\end{eqnarray}
Then, the propagator (\ref{b.4}) becomes 
\begin{eqnarray}
K(x_{f},x_{i};T) &=&\omega \int_{-\infty }^{\infty }d(\beta _{f}-\beta
_{i})\exp \left[ -i\sqrt{\frac{2mV_{0}}{\hbar ^{2}\omega ^{2}}-\frac{1}{4}}%
(\beta _{f}-\beta _{i})\right]   \nonumber \\
&&\ \ \times \int D\alpha D\beta Dp_{\alpha }Dp_{\beta }\exp \left\{ \frac{i%
}{\hbar }\int_{0}^{\omega ^{2}T}\left[ p_{\alpha }\stackrel{.}{\alpha }%
+p_{\beta }\stackrel{.}{\beta }\right. \right.   \nonumber  \label{b.6} \\
&&\ \ \left. \left. -\left( \frac{p_{\alpha }^{2}}{2m}+\frac{p_{\beta }^{2}+%
\frac{\hbar ^{2}}{4}}{2m\cosh ^{2}\alpha }\right) \right] dt\right\} .
\label{b.6}
\end{eqnarray}
Our first task is to bring the $(\beta ,p_{\beta })$-dependent part of the
above path integral into an appropriate form. To do this, we consider the
kernel 
\begin{eqnarray}
K(\beta _{f},\beta _{i};T) &=&\int D\beta Dp_{\beta }\exp \left\{ \frac{i}{%
\hbar }\int_{0}^{\omega ^{2}T}\left[ p_{\beta }\stackrel{.}{\beta }-\frac{%
p_{\beta }^{2}+\frac{\hbar ^{2}}{4}}{2m\cosh ^{2}\alpha }\right] dt\right\} 
\nonumber \\
\  &=&\stackunder{N\rightarrow \infty }{\lim }\int \stackrel{N}{\stackunder{%
j=1}{\prod }}d\beta _{j}\stackrel{N+1}{\stackunder{j=1}{\prod }}\frac{%
dp_{\beta _{j}}}{2\pi \hbar }  \nonumber  \label{b.7} \\
&&\ \ \times \exp \left\{ \frac{i}{\hbar }\stackrel{N+1}{\stackunder{j=1}{%
\sum }}\left[ p_{\beta _{j}}\triangle \beta _{j}-\frac{p_{\beta _{j}}^{2}+%
\frac{\hbar ^{2}}{4}}{2m\cosh ^{2}\alpha _{j}}\omega ^{2}\varepsilon \right]
\right\} .  \label{b.7}
\end{eqnarray}
If we integrate this expression over the variables $\beta _{j}$, we obtain 
\begin{eqnarray}
K(\beta _{f},\beta _{i};T) &=&\stackunder{N\rightarrow \infty }{\lim }\int 
\frac{dp_{\beta _{N+1}}}{2\pi \hbar }\exp \left\{ \frac{i}{\hbar }\left[
\beta _{N+1}p_{\beta _{j}}-\beta _{0}p_{\beta _{1}}\right] \right\} \int 
\stackunder{j=1}{\stackrel{N}{\prod }}dp_{\beta _{j}}  \nonumber  \label{b.8}
\\
&&\ \ \times \delta (p_{\beta _{j}}-p_{\beta _{j+1}})\stackrel{N+1}{%
\stackunder{j=1}{\prod }}\exp \left( -\frac{i}{\hbar }\frac{p_{\beta
_{j}}^{2}+\frac{\hbar ^{2}}{4}}{2m\cosh ^{2}\alpha _{j}}\omega
^{2}\varepsilon \right) .  \label{b.8}
\end{eqnarray}
The $p_{\beta _{j}}$ integrations thereafter give $p_{\beta _{1}}=p_{\beta
_{2}}=...=p_{\beta _{N+1}}=\hbar \lambda $. Hence we have 
\begin{equation}
K(\beta _{f},\beta _{i};T)=\int_{-\infty }^{\infty }\frac{d\lambda }{2\pi }%
\exp \left[ i\lambda \left( \beta _{f}-\beta _{i}\right) \right] \exp \left(
-\frac{i}{\hbar }\int_{0}^{\omega ^{2}T}\frac{\hbar ^{2}}{2m}\frac{\lambda
^{2}+\frac{1}{4}}{\cosh ^{2}\alpha }dt\right) .  \label{b.9}
\end{equation}
Substituting (\ref{b.9}) into (\ref{b.6}) yields the following expression
for the propagator: 
\begin{eqnarray}
K(x_{f},x_{i};T) &=&\omega \int_{-\infty }^{\infty }d(\beta _{f}-\beta
_{i})\exp \left[ -i\sqrt{\frac{2mV_{0}}{\hbar ^{2}\omega ^{2}}-\frac{1}{4}}%
(\beta _{f}-\beta _{i})\right]   \nonumber  \label{b.10} \\
&&\ \ \ \times \int_{-\infty }^{\infty }\frac{d\lambda }{2\pi }\exp \left[
i\lambda \left( \beta _{f}-\beta _{i}\right) \right] K(\alpha _{f},\alpha
_{i};T),  \label{b.10}
\end{eqnarray}
where the $\left( \alpha ,p_{\alpha }\right) $-part is given by\ 
\begin{equation}
K(\alpha _{f},\alpha _{i};T)=\int D\alpha Dp_{\alpha }\exp \left[ \frac{i}{%
\hbar }\int_{0}^{\omega ^{2}T}\left( p_{\alpha }\stackrel{.}{\alpha }-\frac{%
p_{\alpha }^{2}}{2m}-\frac{\hbar ^{2}}{2m}\frac{\lambda ^{2}+\frac{1}{4}}{%
\cosh ^{2}\alpha }\right) dt\right] .  \label{b.11}
\end{equation}
The calculation of this path integral is of course not at all
straightforward. We will now follow Ref.\cite{andand} to convert our
propagator into the Morse potential problem by utilizing a set of tricks. We
start by performing the point canonical transformation defined by 
\begin{equation}
X=\sinh \alpha ,P=\cosh \alpha \text{ }p_{\alpha },  \label{b.12}
\end{equation}
which introduces the effective potential \cite{andand,paksok} 
\begin{equation}
V_{e}=\frac{\hbar ^{2}}{8m}\left( 3-\frac{1}{1+X^{2}}\right) .  \label{b.13}
\end{equation}
The transformed path integral takes the form 
\begin{eqnarray}
K(\alpha _{f},\alpha _{i};T) &=&\left[ \left( 1+X_{f}^{2}\right) \left(
1+X_{i}^{2}\right) \right] ^{\frac{1}{4}}\int DXDP\exp \left\{ \frac{i}{%
\hbar }\int_{0}^{\omega ^{2}T}\left[ P\stackrel{.}{X}\right. \right.  
\nonumber  \label{b.14} \\
&&\ \ \left. \left. -\frac{P^{2}}{2m}\left( 1+X^{2}\right) -\frac{\hbar ^{2}%
}{2m}\frac{\lambda ^{2}}{1+X^{2}}-\frac{3\hbar ^{2}}{8m}\right] dt\right\} .
\label{b.14}
\end{eqnarray}
One can get rid of the $\left( 1+X^{2}\right) ^{-1}$ potential term, by
changing the variable $P$ into $\widetilde{P}$ via 
\begin{equation}
P=\widetilde{P}-\frac{i\hbar \lambda }{1+X^{2}}.  \label{b.15}
\end{equation}
This lets appear a total $X$ derivative which can easily be integrated out.
Then, the path integral (\ref{b.14}) may be rewritten 
\begin{equation}
K(\alpha _{f},\alpha _{i};T)=\left[ \left( 1+X_{f}^{2}\right) \left(
1+X_{i}^{2}\right) \right] ^{\frac{1}{4}}\exp \left[ \lambda (\arctan X\mid
_{X_{i}}^{X_{f}}\right] K(X_{f},X_{i};T),  \label{b.16}
\end{equation}
where 
\begin{eqnarray}
K(X_{f},X_{i};T) &=&\int DXD\widetilde{P}\exp \left\{ \frac{i}{\hbar }%
\int_{0}^{\omega ^{2}T}\left[ \widetilde{P}\stackrel{.}{X}-\frac{\widetilde{P%
}^{2}}{2m}\left( 1+X^{2}\right) \right. \right.   \nonumber  \label{b.17} \\
&&\ \ \left. \left. +\frac{i\hbar \lambda }{m}\widetilde{P}-\frac{3\hbar ^{2}%
}{8m}\right] dt\right\} .  \label{b.17}
\end{eqnarray}
Following Anderson et al \cite{andand} , the $\widetilde{P}^{2}X^{2}$ term
can be absorbed using the trick called uncompleting the square, namely 
\begin{eqnarray}
&&\ \ \exp \left( -\frac{i}{\hbar }\int_{0}^{\omega ^{2}T}\frac{\widetilde{P}%
^{2}}{2m}X^{2}dt\right)   \nonumber  \label{b.18} \\
\  &=&\int_{-\infty }^{\infty }d\xi _{f}\int D\xi \exp \left[ \frac{i}{\hbar 
}\int_{0}^{\omega ^{2}T}\left( \frac{m}{2}\stackrel{.}{\xi }^{2}-\stackrel{.%
}{\xi }\widetilde{P}X+\frac{3\hbar ^{2}}{8m}\right) dt\right] ,  \label{b.18}
\end{eqnarray}
and (\ref{b.17}) takes the form 
\begin{eqnarray}
K(X_{f},X_{i};T) &=&\int_{-\infty }^{\infty }d\xi _{f}\int D\xi \int DXD%
\widetilde{P}\exp \left[ \frac{i}{\hbar }\int_{0}^{\omega ^{2}T}\left( \frac{%
m}{2}\stackrel{.}{\xi }^{2}+\widetilde{P}\stackrel{.}{X}\right. \right.  
\nonumber \\
&&\ \ \left. \left. -\frac{\widetilde{P}^{2}}{2m}-\stackrel{.}{\xi }%
\widetilde{P}X+\frac{i\hbar \lambda }{m}\widetilde{P}\right) dt\right] .
\label{b.19}
\end{eqnarray}
By discretizing the $\widetilde{P}\stackrel{.}{X}$ term and performing the
integrations on the $N$ variables $X_{j\text{ , }}$it appears $N$ Dirac
distributions which can formally be written as 
\begin{equation}
\delta \left( \stackrel{.}{\widetilde{P}}+\stackrel{.}{\xi }\widetilde{P}%
\right) .  \label{b.20}
\end{equation}
Taking into account the boundary conditions $\widetilde{P}(0)=P_{0},\xi (0)=0
$ and $\xi (\omega ^{2}T)=\xi _{f}$ , the solution of the differential
equation in the argument of the Dirac distribution (\ref{b.20}) is \ 
\begin{equation}
\widetilde{P}(t)=P_{0}e^{-\xi (t)}.  \label{b.21}
\end{equation}
The $\widetilde{P}_{j}$ integrations thereafter may be performed and reduced
to an ordinary integration over $P_{0}$. It can also be shown that 
\begin{equation}
\left( \frac{d\widetilde{P}(\omega ^{2}T)}{dP_{0}}\right) ^{\frac{1}{2}}=e^{-%
\frac{\xi _{f}}{2}}.  \label{b.22}
\end{equation}
As a result, we have 
\begin{eqnarray}
K(X_{f},X_{i};T) &=&\int_{-\infty }^{\infty }d\xi _{f}\exp \left( -\frac{\xi
_{f}}{2}\right) \int_{-\infty }^{\infty }\frac{dP_{0}}{2\pi \hbar } 
\nonumber  \label{b.19} \\
&&\ \ \times \exp \left[ \frac{i}{\hbar }P_{0}\left( e^{-\xi
_{f}}X_{f}-X_{i}\right) \right] K_{M}(\xi _{f},0;T),  \label{b.23}
\end{eqnarray}
with 
\begin{equation}
K_{M}(\xi _{f},0;T)=\int D\xi \exp \left[ \frac{i}{\hbar }\int_{0}^{\omega
^{2}T}\left( \frac{m}{2}\stackrel{.}{\xi }^{2}-\frac{P_{0}^{2}}{2m}e^{-2\xi
}+\frac{i\hbar \lambda }{m}P_{0}e^{-\xi }\right) dt\right] .  \label{b.24}
\end{equation}
This kernel describes the path integral for a particle to go from a position 
$\xi _{i}=0$ at time $t_{i}=0$ to $\xi _{f}$ at time $t_{f}=\omega ^{2}T$ ,
in the Morse-type potential 
\begin{equation}
V(\xi )=\frac{P_{0}^{2}}{2m}\left( e^{-2\xi }-\frac{2i\hbar \lambda }{P_{0}}%
e^{-\xi }\right) .  \label{b.25}
\end{equation}
For $\frac{2i\hbar \mu }{P_{0}}$ real and positive, there exist treatments
by path integral techniques \cite{khan,klei,paksok,cainowi,duru} ,operator
methods \cite{nietsim} and algebraic approach \cite{chegueham} . If one
substitutes $A=\frac{P_{0}^{2}}{2m},B=\frac{i\hbar \lambda }{m}%
P_{0},a=1,M=4m,$ $s=\frac{2m}{P_{0}}S$ in Ref.\cite{chegueham} , one obtains
\ 
\begin{eqnarray}
K_{M}(\xi _{f},0;T) &=&\frac{2m}{\hbar }\int_{0}^{\infty }\frac{e^{-2\lambda
S}}{i\sin (S)}dS\int_{-\infty }^{\infty }\frac{dE}{2\pi \hbar }\exp \left( -%
\frac{i}{\hbar }E\omega ^{2}T\right)   \nonumber \\
&&\ \times \exp \left[ \frac{i}{\hbar }P_{0}\left( e^{-\xi _{f}}+1\right)
\cot (S)\right]   \nonumber  \label{b.26} \\
&&\ \times I_{2\sqrt{-2mE/\hbar ^{2}}}\left( \frac{2P_{0}e^{-\frac{\xi _{f}}{%
2}}}{i\hbar \sin (S)}\right) .  \label{b.26}
\end{eqnarray}
With the results (\ref{b.26}), (\ref{b.23}) and (\ref{b.16}) the full
propagator (\ref{b.10}) for the potential under consideration may be put
into the form 
\begin{eqnarray}
K(x_{f},x_{i};T) &=&\frac{2m\omega }{\hbar }\left[ \left( 1+X_{f}^{2}\right)
\left( 1+X^{2}\right) \right] ^{\frac{1}{4}}\exp \left[ \lambda (\arctan
X\mid _{X_{i}}^{X_{f}}\right]   \nonumber \\
&&\ \times \int_{-\infty }^{\infty }d(\beta _{f}-\beta _{i})\exp \left[ -i%
\sqrt{\frac{2mV_{0}}{\hbar ^{2}\omega ^{2}}-\frac{1}{4}}\left( \beta
_{f}-\beta _{i}\right) \right]   \nonumber \\
&&\ \times \int_{-\infty }^{\infty }\frac{d\lambda }{2\pi }\exp \left[
i\lambda \left( \beta _{f}-\beta _{i}\right) \right] \int_{-\infty }^{\infty
}d\xi _{f}\exp \left( -\frac{\xi _{f}}{2}\right) \int_{-\infty }^{\infty }%
\frac{dP_{0}}{2\pi \hbar }  \nonumber \\
&&\ \times \exp \left[ \frac{i}{\hbar }P_{0}\left( e^{-\xi
_{f}}X_{f}-X_{i}\right) \right] \int_{0}^{\infty }\frac{e^{-2\lambda S}}{%
i\sin (S)}dS\int_{-\infty }^{\infty }\frac{dE}{2\pi \hbar }  \nonumber \\
&&\ \times \exp \left( -\frac{i}{\hbar }E\omega ^{2}T\right) \exp \left[ 
\frac{i}{\hbar }P_{0}\cot (S)\left( e^{-\xi _{f}}+1\right) \right]  
\nonumber \\
&&\ \times I_{2\sqrt{-2mE/\hbar ^{2}}}\left( \frac{2P_{0}e^{-\frac{\xi _{f}}{%
2}}}{i\hbar \sin (S)}\right) ,  \label{b.27}
\end{eqnarray}
where $X_{f}=\sinh \alpha _{f}$ and $X_{i}=\sinh \alpha _{i}$.

\section{Evaluation of the propagator}

In order to evaluate the multiple integral (\ref{b.27}), we start by
employing \ the integral representation for the modified Bessel function

\begin{equation}
-2\pi iI_{\mu }(-2iaz)=z^{2\mu }\int_{-\infty }^{\infty }duu^{-(1+2\mu
)}\exp \left[ -ia\left( u+\frac{z^{2}}{u}\right) \right] ,  \label{b.28}
\end{equation}
valid for $\arg (a)=0$ and $Re(2\mu )\succ -1$, changing variables $%
q=e^{-\xi _{f}},\widetilde{q}=q/u^{2}$ and $\widetilde{P}_{0}=P_{0}/\sin (S)$
and carrying out the integration over $\widetilde{P}_{0}$. We then obtain 
\begin{eqnarray}
K(x_{f},x_{i};T) &=&\frac{2m\omega }{\hbar }\left[ \left( 1+X_{f}^{2}\right)
\left( 1+X^{2}\right) \right] ^{\frac{1}{4}}\exp \left[ \lambda (\arctan
X\mid _{X_{i}}^{X_{f}}\right]   \nonumber \\
&&\times \int_{-\infty }^{\infty }d(\beta _{f}-\beta _{i})\exp \left[ -i%
\sqrt{\frac{2mV_{0}}{\hbar ^{2}\omega ^{2}}-\frac{1}{4}}\left( \beta
_{f}-\beta _{i}\right) \right]   \nonumber \\
&&\ \times \int_{-\infty }^{\infty }\frac{d\lambda }{2\pi }\exp \left[
i\lambda \left( \beta _{f}-\beta _{i}\right) \right] \int_{-\infty }^{\infty
}\frac{dE}{2\pi \hbar }  \nonumber \\
&&\ \times \exp \left( -\frac{i}{\hbar }E\omega ^{2}T\right)
\int_{0}^{\infty }dSe^{-2\lambda S}\int_{0}^{\infty }\widetilde{q}^{\mu -%
\frac{1}{2}}d\widetilde{q}  \nonumber \\
&&\times \int_{-\infty }^{\infty }\delta \left( \widetilde{q}\frak{B}%
u^{2}-u(1+\widetilde{q})+\frak{A}\right) du,  \label{b.29}
\end{eqnarray}
where we have set 
\begin{equation}
\frak{A}=\cos (S)+X_{f}\sin (S)\text{ , \quad }\frak{B}=\cos (S)-X_{i}\sin
(S),  \label{b.30}
\end{equation}
and 
\begin{equation}
\mu =\sqrt{-\frac{2mE}{\hbar ^{2}}}=\pm \frac{ik}{\omega }.  \label{b.31}
\end{equation}
Next, the integration over $u$ yields 
\begin{equation}
\int_{-\infty }^{\infty }\delta \left( \widetilde{q}\frak{B}u^{2}-u(1+%
\widetilde{q})+\frak{A}\right) du=\frac{2}{\sqrt{\widetilde{q}^{2}+2(1-2%
\frak{AB})\widetilde{q}+1}}.  \label{b.32}
\end{equation}
If we let 
\begin{equation}
S=\widetilde{S}+\frac{1}{2}\left( \arctan X_{f}-\arctan X_{i}\right)
,\triangle =-\left( \arctan X_{f}+\arctan X_{i}\right) ,  \label{b.33}
\end{equation}
and noticing that 
\begin{equation}
\cos (S)\mp X\sin (S)=\sqrt{1+X^{2}}\cos \left( S\pm \arctan X\right) ,
\label{b.34}
\end{equation}
the factor $\left( 1-2\frak{AB}\right) $ is amenable to the form 
\begin{equation}
\left( 1-2\frak{AB}\right) =\sinh \alpha _{f}\sinh \alpha _{i}-\cosh \alpha
_{f}\cosh \alpha _{i}\cos (2\widetilde{S}).  \label{b.35}
\end{equation}
The integrations over the variables $\widetilde{q},$ $\lambda $ and $%
\widetilde{S}$ can be easily performed using the integral ( see formula $%
3.252.11$ $p.297$ in \cite{gradryz} ) 
\begin{eqnarray}
&&\ \int_{0}^{\infty }\left( 1+2\beta x+x^{2}\right) ^{\mu -\frac{1}{2}%
}x^{^{-\nu -1}}dx  \nonumber \\
\  &=&2^{-\mu }\left( \beta ^{2}-1\right) ^{\frac{\mu }{2}}\Gamma (1-\mu
)B(\nu -2\mu +1,-\nu )P_{\nu -\mu }^{\mu }(\beta ),  \label{b.36}
\end{eqnarray}
valid for $Re(\nu )\prec 0,Re(2\mu -\nu )\prec 1,$ and $\left| \arg (\beta
\pm 1)\right| \prec \pi ,$ where $B(x,y)=\frac{\Gamma (x)\Gamma (y)}{\Gamma
(x+y)}.$

Taking into account the symmetry property of the general associated Legendre
functions (see formula $8.731.5$ $p.1005$ in \cite{gradryz} ) 
\begin{equation}
P_{-\frac 12-i\lambda }^\mu (x)=P_{-\frac 12+i\lambda }^\mu (x),
\label{b.37}
\end{equation}
and the well-known relation ( see formula $8.334.2$ $p.937$ in \cite{gradryz}%
) 
\begin{equation}
\Gamma (\frac 12-x)\Gamma (\frac 12+x)=\frac \pi {\cos \pi x},  \label{b.38}
\end{equation}
we obtain\ 
\begin{eqnarray}
K(x_f,x_i;T) &=&\frac{\sqrt{\cosh \alpha _f\cosh \alpha _i}}{\pi \omega ^2}%
\int_{-\infty }^\infty \frac{kdk}{\cosh \left( \frac{\pi k}\omega \right) }%
\exp \left( -\frac i\hbar \frac{\hbar ^2k^2}{2m}T\right)  \nonumber \\
&&\ \ \times \int_{-\infty }^\infty d(\beta _f-\beta _i)\exp \left[ -i\sqrt{%
\frac{2mV_0}{\hbar ^2\omega ^2}-\frac 14}\left( \beta _f-\beta _i\right) %
\right]  \nonumber  \label{b.39} \\
&&\ \times P_{-\frac 12+\frac{ik}\omega }(-Z),  \label{b.39}
\end{eqnarray}
where 
\begin{equation}
Z=\cosh \alpha _f\cosh \alpha _i\cosh (\beta _f-\beta _i)-\sinh \alpha
_f\sinh \alpha _i.  \label{b.40}
\end{equation}
The variables $\alpha _f,\alpha _i,\beta _f,$ and $\beta _i$ in $(39)$ can
be separated with the help of the addition theorem derived in Ref.\cite
{ahmedu} 
\begin{equation}
P_{-\frac 12+\frac{ik}\omega }(-Z)=\frac 1{4\pi }\stackrel{1}{\stackunder{%
\varepsilon =0}{\sum }}\int_{-\infty }^\infty d\lambda e^{-i\lambda (\beta
_f-\beta _i)}d_{\lambda ,0}^{-\frac 12+\frac{ik}\omega ,\varepsilon }(\alpha
_f+i\pi )\left( d_{\lambda ,0}^{-\frac 12+\frac{ik}\omega ,\varepsilon
}(\alpha _i)\right) ^{*},  \label{b.41}
\end{equation}
where 
\begin{equation}
d_{\lambda ,0}^{-\frac 12+\frac{ik}\omega ,\varepsilon }(\alpha )=\frac
1{2\pi }\int_{-\infty }^\infty d\beta \exp (-i\lambda \beta )\left[ \cosh
\beta \cosh \alpha +(-1)^\varepsilon \sinh \alpha \right] ^{-\frac 12+\frac{%
ik}\omega },  \label{b.42}
\end{equation}
which may be evaluated by using the integral ( see formula $8.713.3$ $p.1001$
in \cite{gradryz} ) 
\begin{equation}
P_\nu ^{-\gamma }(z)=\sqrt{\frac 2\pi }\frac{\Gamma \left( \gamma +\frac
12\right) \left( z^2-1\right) ^{\frac \gamma 2}}{\Gamma (\nu +\gamma
+1)\Gamma (\gamma -\nu )}\int_0^\infty \frac{\cosh \left[ \left( \nu +\frac
12\right) t\right] dt}{\left( z+\cosh t\right) ^{\gamma +\frac 12}},
\label{b.43}
\end{equation}
valid for $Re(z)\succ -1,\left| \arg (z\pm 1)\right| \prec \pi ,Re(\nu
+\gamma )\succ -1,Re(\nu -\gamma )\succ 0,$ and we have\ 
\begin{equation}
d_{\lambda ,0}^{-\frac 12+\frac{ik}\omega ,\varepsilon }(\alpha )=\frac{%
\Gamma \left( \frac 12+i\lambda -\frac{ik}\omega \right) \Gamma \left( \frac
12-i\lambda -\frac{ik}\omega \right) }{\sqrt{2\pi }\Gamma \left( \frac 12-%
\frac{ik}\omega \right) }\frac{e^{-\frac{\pi k}{2\omega }}}{\sqrt{\cosh
\alpha }}P_{-\frac 12+i\lambda }^{\frac{ik}\omega }\left( (-1)^\varepsilon
\tanh \alpha \right) .  \label{b.44}
\end{equation}
Inserting (\ref{b.44}) and (\ref{b.41}) in (\ref{b.39}) and integrating over
the variable $(\beta _f-\beta _i)$, we arrive at 
\begin{eqnarray}
K(x_f,x_i;T) &=&-\frac 1{2\omega }\int_{-\infty }^\infty \frac{kdk}{\left|
\sin \left[ \pi \left( \nu -\frac{ik}\omega \right) \right] \right| ^2}e^{-%
\frac{\pi k}\omega }\stackrel{1}{\stackunder{\varepsilon =0}{\sum }}\left(
P_\nu ^{\frac{ik}\omega }((-1)^\varepsilon \tanh \alpha _f)\right.  \nonumber
\\
&&\ \times \left. P_\nu ^{-\frac{ik}\omega }((-1)^\varepsilon \tanh \alpha
_i)\right) \exp \left( -\frac i\hbar \frac{\hbar ^2k^2}{2m}T\right) ,
\label{b.45}
\end{eqnarray}
where 
\begin{equation}
\nu =-\frac 12+i\lambda =\frac 12\left( -1+\sqrt{1-\frac{8mV_0}{\hbar
^2\omega ^2}}\right) .  \label{b.46}
\end{equation}
In order to find the wave functions, we set

\begin{equation}
\Psi _1(y)=P_\nu ^{-\mu }(y),\quad \Psi _2(y)=P_\nu ^{-\mu }(-y).
\label{b.47}
\end{equation}
The parameter $\nu $ can be real or complex. In the complex case, we have 
\begin{equation}
1+\nu ^{*}=-\nu .  \label{b.48}
\end{equation}
Owing to (\ref{b.37}), it is obvious that 
\begin{equation}
\Psi _1^{*}(y)=P_\nu ^\mu (y),\quad \Psi _2^{*}(y)=P_\nu ^\mu (-y),
\label{b.49}
\end{equation}
and thanks to the relations ( see formulas $8.737.1$ and $8.737.2$ $p.1006$
in \cite{gradryz} )

\begin{equation}
\left\{ 
\begin{array}{c}
P_{\overline{\nu }}^{-\mu }(y)=\frac{\Gamma (\nu -\mu +1)}{\Gamma (\nu +\mu
+1)}\left[ \cos (\pi \mu )P_\nu ^\mu (y)-\frac 2\pi \sin (\pi \mu )Q_\nu
^\mu (y)\right] , \\ 
P_\nu ^\mu (-y)=\cos \left[ \pi (\nu +\mu )\right] P_\nu ^\mu (y)-\frac 2\pi
\sin \left[ \pi (\nu +\mu )\right] Q_\nu ^\mu (y),
\end{array}
\right.  \label{b.50}
\end{equation}
\ it is easy to make up the following equations 
\begin{equation}
\left\{ 
\begin{array}{c}
\Psi _1(y)=a\Psi _1^{*}(y)+b\Psi _2^{*}(y), \\ 
\Psi _2^{*}(y)=a^{*}\Psi _2(y)+b^{*}\Psi _1(y),
\end{array}
\right.  \label{b.51}
\end{equation}
where 
\begin{equation}
a=\frac{\Gamma (\nu -\mu +1)\sin (\pi \nu )}{\Gamma (\nu +\mu +1)\sin \left[
\pi (\nu +\mu )\right] }\text{ , }b=\frac{\Gamma (\nu -\mu +1)\sin (\pi \mu )%
}{\Gamma (\nu +\mu +1)\sin \left[ \pi (\nu +\mu )\right] }.  \label{b.52}
\end{equation}
Expression (\ref{b.45}) of the propagator is then rewritten

\begin{eqnarray}
K(x_f,x_i;T) &=&-\frac 1{2\omega }\int_{-\infty }^\infty \frac{kdk}{\left|
\sin \left[ \pi \left( \nu -\frac{ik}\omega \right) \right] \right| ^2}e^{-%
\frac{\pi k}\omega }\left( \Psi _1(y_f)\Psi _1^{*}(y_i)\right.  \nonumber \\
&&\left. +\Psi _2(y_f)\Psi _2^{*}(y_i)\right) \exp \left( -\frac i\hbar 
\frac{\hbar ^2k^2}{2m}T\right) .  \label{b.53}
\end{eqnarray}
Next by changing $k$ into $(-k)$ in the interval of integration $\left]
-\infty ,0\right] ,$and substituting $\Psi _j(y)$ and $\Psi _j^{*}(y)$ by
their expressions (\ref{b.51}), we get 
\begin{eqnarray}
K(x_f,x_i;T) &=&\frac m{2\hbar ^2\omega }\int_0^\infty dE_k\frac{\sinh
\left( \frac{\pi k}\omega \right) }{\left| \sin \left[ \pi \left( \nu -\frac{%
ik}\omega \right) \right] \right| ^2}\exp \left( -\frac i\hbar E_kT\right) 
\nonumber \\
&&\ \times \left\{ \left( \left| a\right| ^2+\left| b\right| ^2\right)
\left( \Psi _1(y_f)\Psi _1^{*}(y_i)+\Psi _2(y_f)\Psi _2^{*}(y_i)\right)
\right.  \nonumber \\
&&\left. +\left( ab^{*}+a^{*}b\right) \left( \Psi _1(y_f)\Psi
_2^{*}(y_i)+\Psi _2(y_f)\Psi _1^{*}(y_i)\right) \right\} .  \label{b.54}
\end{eqnarray}
Now it is easy to get assured that

\begin{equation}
\left| a\right| ^2+\left| b\right| ^2=1\text{ , }\left( ab^{*}+a^{*}b\right)
=0,\forall \nu .  \label{b.55}
\end{equation}
Finally, the expression of the propagator is given by\ 
\begin{eqnarray}
K(x_f,x_i;T) &=&\frac m{2\hbar ^2\omega }\int_0^\infty dE_k\frac{\sinh
\left( \frac{\pi k}\omega \right) }{\left| \sin \left[ \pi \left( \nu -\frac{%
ik}\omega \right) \right] \right| ^2}  \nonumber \\
&&\ \times \left\{ \Psi _1(y_f)\Psi _1^{*}(y_i)+\Psi _2(y_f)\Psi
_2^{*}(y_i)\right\} \exp \left( -\frac i\hbar E_kT\right) .  \label{b.56}
\end{eqnarray}
The suitably normalized wave functions, corresponding to the energy\newline
$E_k=\frac{\hbar ^2k^2}{2m},$ are readily obtained to be 
\begin{equation}
\Psi _{\stackrel{\rightarrow }{\leftarrow }}(y)=\sqrt{\frac m{2\hbar
^2\omega }}\frac{\left[ \sinh \left( \frac{\pi k}\omega \right) \right]
^{\frac 12}}{\left| \sin \left[ \pi \left( \nu -\frac{ik}\omega \right) %
\right] \right| }P_\nu ^{\frac{ik}\omega }(\pm \tanh \alpha ),  \label{b.57}
\end{equation}
where we have used the standard convention according to which the arrow in
the wave function indicates the way in which the waves are propagating.

Taking into account the well-known link between the general associated
Legendre functions and the hypergeometric functions ( see formula $8.704$ $%
p.999$ in \cite{gradryz} ) 
\begin{equation}
P_\nu ^\mu (x)=\frac 1{\Gamma (1-\mu )}\left[ \frac{1+x}{1-x}\right] ^{\frac
\mu 2}F\left( -\nu ,\nu +1;1-\mu ;\frac{1-x}2\right) ,  \label{b.58}
\end{equation}
and using the relation ( see formula $9.131.1$ $p.1043$ in \cite{gradryz} ) 
\begin{equation}
F(\alpha ,\beta ;\gamma ;z)=(1-z)^{\gamma -\alpha -\beta }F(\gamma -\alpha
,\gamma -\beta ;\gamma ;z),  \label{b.59}
\end{equation}
the wave functions are written\ 
\begin{eqnarray}
\Psi _{\stackrel{\rightarrow }{\leftarrow }}(y) &=&\sqrt{\frac m{2\hbar
^2\omega }}\frac{\left[ \sinh \left( \frac{\pi k}\omega \right) \right]
^{\frac 12}}{\left| \sin \left[ \pi \left( \nu -\frac{ik}\omega \right) %
\right] \right| }\frac 1{\Gamma (1-\frac{ik}\omega )}\left[ \frac{1-\tanh
^2\alpha }4\right] ^{-\frac{ik}{2\omega }}  \nonumber \\
&&\ \times F\left( 1+\nu -\frac{ik}\omega ,-\nu -\frac{ik}\omega ;1-\frac{ik}%
\omega ;\frac{1\mp \tanh \alpha }2\right) .  \label{b.60}
\end{eqnarray}

\section{Scattering function}

The sum of the reflection and transmission amplitudes $R(k)$ and $T(k)$ or
else the scattering function $S(k)$ is independent of the way in which the
waves are propagating. As an example, we can consider the case described by $%
\Psi _{\rightarrow }(y)$ in order to evaluate $S(k)$.

Using the identities for the hypergeometric functions ( see formula $9.131.2$
$p.1043$ in \cite{gradryz} ) 
\begin{equation}
F\left( \alpha ,\beta ;\gamma ;\frac{1-y}2\right) \rightarrow \left\{ 
\begin{array}{c}
1,\quad y\rightarrow 1, \\ 
\frac{\Gamma (\gamma )\Gamma (\gamma -\alpha -\beta )}{\Gamma (\gamma
-\alpha )\Gamma (\gamma -\beta )}+\left( \frac{1+y}2\right) ^{\gamma -\alpha
-\beta }\frac{\Gamma (\gamma )\Gamma (\alpha +\beta -\gamma )}{\Gamma
(\alpha )\Gamma (\beta )},y\rightarrow -1,
\end{array}
\right.  \label{b.61}
\end{equation}
we obtain 
\begin{eqnarray}
\Psi _{\rightarrow }(y) &\rightarrow &\sqrt{\frac m{2\hbar ^2\omega }}\frac{%
\left[ \sinh \left( \frac{\pi k}\omega \right) \right] ^{\frac 12}}{\left|
\sin \left[ \pi \left( \nu -\frac{ik}\omega \right) \right] \right| } 
\nonumber \\
&&\ \times \left\{ 
\begin{array}{c}
\frac 1{\Gamma (1-\frac{ik}\omega )}e^{ikx},\quad x\rightarrow +\infty \\ 
\frac{\Gamma \left( -\frac{ik}\omega \right) }{\Gamma \left( 1+\nu -\frac{ik}%
\omega \right) \Gamma \left( -\nu -\frac{ik}\omega \right) }e^{ikx}+\frac{%
\Gamma \left( \frac{ik}\omega \right) }{\Gamma \left( 1+\nu \right) \Gamma
\left( -\nu \right) }e^{-ikx},x\rightarrow -\infty .
\end{array}
\right.  \label{b.62}
\end{eqnarray}
Identifying these asymptotic behaviors with the boundary conditions 
\begin{equation}
\Psi _{\rightarrow }(y)\rightarrow \left\{ 
\begin{array}{c}
T(k)e^{ikx},\quad x\rightarrow +\infty \\ 
e^{ikx}+R(k)e^{-ikx},\quad x\rightarrow -\infty ,
\end{array}
\right.  \label{b.63}
\end{equation}
the amplitudes $T(k)$ and $R(k)$ of the transmitted and reflected beams are
evaluated to be 
\begin{equation}
\left\{ 
\begin{array}{c}
T(k)=\frac{\Gamma \left( 1+\nu -\frac{ik}\omega \right) \Gamma \left( -\nu -%
\frac{ik}\omega \right) }{\Gamma \left( 1-\frac{ik}\omega \right) \Gamma
\left( -\frac{ik}\omega \right) }, \\ 
R(k)=\frac{\Gamma \left( 1+\nu -\frac{ik}\omega \right) \Gamma \left( -\nu -%
\frac{ik}\omega \right) \Gamma \left( \frac{ik}\omega \right) }{\Gamma
\left( 1+\nu \right) \Gamma \left( -\nu \right) \Gamma \left( -\frac{ik}%
\omega \right) },
\end{array}
\right.  \label{b.64}
\end{equation}
from which the expression of the scattering function is 
\begin{equation}
S(k)=T(k)+R(k)=\frac{\Gamma \left( \frac{ik}\omega \right) \Gamma \left(
-\nu -\frac{ik}\omega \right) \cos \left[ \left( \frac \pi 2\right) \left(
\nu +\frac{ik}\omega \right) \right] }{\Gamma \left( -\frac{ik}\omega
\right) \Gamma \left( -\nu +\frac{ik}\omega \right) \cos \left[ \left( \frac
\pi 2\right) \left( \nu -\frac{ik}\omega \right) \right] },  \label{b.65}
\end{equation}
and is clearly unitary 
\begin{equation}
S^{*}(k)=S^{-1}(k).  \label{b.66}
\end{equation}
Note that the analysis of the asymptotic behavior of $\Psi _{\leftarrow }(y)$
with the boundary conditions 
\begin{equation}
\Psi _{\leftarrow }(y)\rightarrow \left\{ 
\begin{array}{c}
T(k)e^{-ikx},\quad x\rightarrow -\infty \\ 
e^{-ikx}+R(k)e^{ikx},\quad x\rightarrow +\infty ,
\end{array}
\right.  \label{b.67}
\end{equation}
leads to the same expression of the scattering function.

The exact expressions for $T(k)$ and $R(k)$ in Eqs. (\ref{b.64}) differ
significantly from the ones deduced from the solution \cite{ahmedu} obtained
through path integration over $SO(1,2)$.

\section{Particular cases}

We consider now two limiting cases.

\subsection{The free particle}

We have to check whether for $V_0\rightarrow 0$, we obtain the well-known
coefficients $T(k)$ and $R(k)$ relative to a free particle.

From Eq. (\ref{b.46}), we notice that 
\begin{equation}
\stackunder{V_0\rightarrow 0}{\nu }\rightarrow 0.  \label{b.68}
\end{equation}
Thus, we obtain from Eqs. (\ref{b.64})$:$%
\begin{equation}
\stackunder{\nu \rightarrow 0}{\lim T(k)}=1\text{ , }\stackunder{\nu
\rightarrow 0}{\quad \lim R(k)}=0.  \label{b.69}
\end{equation}
These equations obviously prove that the particle is free.

\subsection{The $\protect\delta -$function barrier}

By letting $V_0\rightarrow \infty $ and $\omega \rightarrow \infty $ such
that $\frac{2V_0}\omega \rightarrow \frac{\hbar ^2}{2m}g=const.,$ we have 
\begin{equation}
\int_{-\infty }^\infty \frac{V_0}{\cosh ^2(\omega x)}dx=\frac{\hbar ^2}{2m}g,
\label{b.70}
\end{equation}
hence the potential (\ref{b.1}) becomes 
\begin{equation}
V(x)=\frac{\hbar ^2}{2m}g\delta (x).  \label{b.71}
\end{equation}
In this case 
\begin{equation}
\stackunder{(V_0\rightarrow \infty ,\omega \rightarrow \infty )}{\nu }%
\rightarrow -\frac g{2\omega }.  \label{b.72}
\end{equation}
Thus, 
\begin{equation}
\stackunder{(V_0\rightarrow \infty ,\omega \rightarrow \infty )}{\lim }T(k)=%
\frac{2k}{2k+ig},\quad \text{ }\stackunder{(V_0\rightarrow \infty ,\omega
\rightarrow \infty )}{\lim }R(k)=\frac{-ig}{2k+ig}.  \label{b.73}
\end{equation}
One can then easily see that if one substitutes $g=-\gamma $, one obtains
the amplitudes of transmitted and reflected beams for the attractive $\delta
-$function barrier given in Ref \cite{gott} .

\section{Conclusion}

The main result of this article concerns the construction of the propagator
relative to the potential barrier $V=\cosh ^{-2}(\omega x)$. In carrying out
the path integration, the method of delta functionals has enabled us to
connect the potential barrier to the Morse potential problem. The evaluation
of the multiple integral representation of the propagator is facilitated by
using the technique of dimensional extension.

It seems to us that this method is a valuable alternative to the treatment
of this potential by path integration over the $SO(1,2)$ manifold.

We have also extracted the correctly normalized wave functions. The
scattering function is obtained from their asymptotic behavior . Two
limiting cases are considered. They constitute a consistency check for the
correctness of our results.


\begin{thebibliography}{99}
\bibitem{khan}  D.C. Khandekar and S.V. Lawande, Phys. Rep. \textbf{137} (
1986 ) 115; D.C. Khandekar, S.V. Lawande and K.V. Bhagwat, Path Integral
Methods and Their Applications ( World Scientific, Singapore, 1993 ).

\bibitem{klei}  H. Kleinert, Path Integrals in Quantum Mechanics, Statistics
and Polymer Physics ( World Scientific, Singapore, 1990 ).

\bibitem{inokurger}  A. Inomata, H. Kuratsuji and C.C. Gerry, Path Integrals
and Coherent States of $SU(2)$ and $SU(1,1)$ ( World Scientific, Singapore,
1992 ).

\bibitem{grostei}  C. Grosche and F. Steiner, Handbook of Feynman Path
Integrals ( Springer Tracts in Modern Physics, 1998 ).

\bibitem{durklei}  I.H. Duru and H. Kleinert, Fortschr. Phys. \textbf{30} 
\textbf{( }1982 ) 401.

\bibitem{kleimus}  H. Kleinert and I. Mustapic, J. Math. Phys. \textbf{33} (
1992 ) 643.

\bibitem{chegueleha1}  L. Chetouani, L. Guechi, A. Lecheheb and T. F.
Hammann, J. Math. Phys. \textbf{34} ( 1993 ) 1257.

\bibitem{chegueleha2}  L. Chetouani, L. Guechi, A. Lecheheb and T.F.
Hammann, Czech. J. Phys. \textbf{45} ( 1995 ) 699.

\bibitem{cheguelehame}  L. Chetouani, L. Guechi, A. Lecheheb , T. F. Hammann
and A. Messouber, Nuovo Cimento \textbf{B113} ( 1998 ) 81.

\bibitem{gros1}  C. Grosche, J. Phys. A: Math. Gen. \textbf{29} ( 1996 ) 365.

\bibitem{junboh}  G. Junker and M. B\"{o}hm, Phys. Lett. \textbf{A117} (
1986 ) 375.

\bibitem{bohjun}  M. B\"{o}hm and G. Junker, J. Math. Phys. \textbf{28} (
1987 ) 1978.

\bibitem{barinojun}  A.O. Barut,A. Inomata and G. Junker, J. Phys. A: Math.
Gen. \textbf{20},

( 1987 ) 6271; \textbf{23}, ( 1990 ) 1179.

\bibitem{bouchegueha}  T. Boudjedaa, L. Chetouani, L. Guechi and T. F.
Hammann, J. Math. Phys. \textbf{32} ( 1991 ) 441.

\bibitem{gros2}  C. Grosche, J. Math. Phys. \textbf{32} ( 1991 ) 1984.

\bibitem{ahmedu}  H. Ahmedov and I.H. Duru, J. Phys. A: Math. Gen. \textbf{30%
} ( 1997 ) 173.

\bibitem{andand}  A. Anderson and S.B. Anderson, Ann.Phys. ( N.Y ) \textbf{%
199} ( 1990 ) 155.

\bibitem{peakinolanino}  D. Peak and A. Inomata, J. Math. Phys. \textbf{10}
( 1969 ) 1422; W. Langguth and A. Inomata, J. Math. Phys. \textbf{20} ( 1979
) 499.

\bibitem{paksok}  N.K. Pak and I. S\"{o}kmen, Phys. Rev. \textbf{A 30} (
1984 ) 1629.

\bibitem{cainowi}  P.Y. Cai, A. Inomata and R. Wilson, Phys. Lett. \textbf{%
A96} ( 1983 ) 117.

\bibitem{duru}  I.H. Duru, Phys. Rev. \textbf{D28} ( 1983 ) 2689.

\bibitem{nietsim}  M.M. Nieto and L. M.Jr. Simmons, Phys. Rev. \textbf{D20}
( 1979 ) 2689.

\bibitem{chegueham}  L. Chetouani, L. Guechi and T.F. Hammann, Helv. Phys.
Acta \textbf{65}

( 1992 ) 1069.

\bibitem{gradryz}  I.S. Gradshtein and I.M. Ryzhik, Tables of Integrals,
Series and Products ( Academic Press, NewYork 1965 ).

\bibitem{gott}  K. Gottfried, Quantum Mechanics, Vol. I : Fundamentals (
W.A. Benjamin, Inc. London 1966 ).
\end{thebibliography}
\end{document}